\documentclass[a4paper,11pt]{article}
\usepackage{pos}
\usepackage{hyperref}
\newcommand{\U}[1]{\ensuremath{\mathrm{\ #1}}}
\newcommand{\UU}[2]{\ensuremath{\mathrm{\ #1^{#2}}}}

\title{VERITAS observations of the Be/X-ray binary system LS V +44 17 during a major outburst}
 \ShortTitle{VERITAS observations of LS V +44 17}

\author*[a]{Jamie Holder}

\affiliation[a]{Department of Physics and Astronomy and the Bartol Research Institute,\\ University of Delaware, Newark, DE 19713, USA.}

\onbehalf{for the VERITAS collaboration} 

\emailAdd{jholder@udel.edu}

\abstract{The Be/X-ray binary system LS V +44 17 (RX J0440.9+4431) is a potential member of the rare class of gamma-ray binaries. The system is comprised of a Be star and a neutron star companion with an orbital period of 150 days. In December of 2022, MAXI detected an X-ray outburst from the source, which peaked in early January before declining and then rebrightening. During the second peak, the flux exceeded 1 Crab in the 15-50 keV range, and exhibited a pulsed emission component with a pulse period of 208 seconds. VERITAS observations were conducted close to the peak of the second outburst, from January 24 to January 27, 2023 . We report here on the search for very high energy (VHE) gamma-ray emission in these data. }

\ConferenceLogo{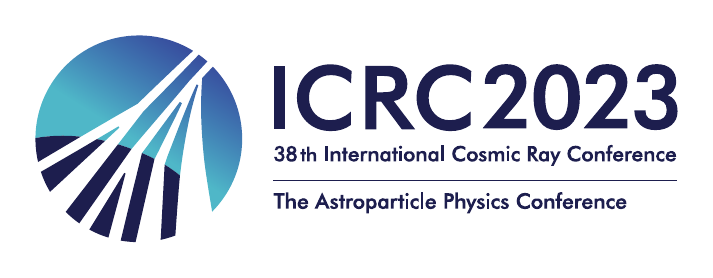}

\FullConference{%
38th International Cosmic Ray Conference (ICRC2023)\\
  26 July - 3 August, 2023\\
  Nagoya, Japan}

\begin{document}
\maketitle

\section{Introduction}
The class of gamma-ray binary systems encompasses a small group of astrophysical objects with a diverse set of properties and emission behaviour. Members are usually defined as binary systems that comprise a compact object (black hole or neutron star) and a stellar companion (typically an O-star,  or a Be-star with a circumstellar disk), in which the peak of the spectral energy distribution lies above $1\U{MeV}$ \cite{2013AnARv..21...64D}. At these high energies, the emission is studied either by space-based gamma-ray telescopes, such as Fermi-LAT (from around 0.1 to $100\U{GeV}$), or by ground-based facilities including imaging atmospheric Cherenkov telescopes and particle detector arrays (above $\sim100\U{GeV}$). In at least two systems, PSR B1259-63/LS 2883 \cite{1259} and PSR J2032+4127/MT91 213 \cite{2032}, the TeV emission is known to be powered by the interaction between an energetic pulsar spin-down wind and the wind and/or disk of the stellar companion. For the remaining gamma-ray binaries the nature of the compact object is not known, but the pulsar-wind model provides a plausible explanation. 

Pulsar-wind powered systems contrast with accreting X-ray pulsars, a class of X-ray binary (XRB) systems in which the emission is powered by accretion of the stellar wind onto a neutron star. In accreting systems with Be-star companions (BeXRBs), the X-ray emission is transient, and is characterized by outbursts classed as Type I (normal) or Type II (giant). Giant outbursts are rare, extremely luminous events ($10^{37}$--$10^{38}\U{erg}\UU{s}{-1}$), which can last for multiple binary orbits. Despite intensive searches over a range of source states, no TeV emission has been consistently detected from any accreting X-ray binary systems. The detection of such emission would likely require the development or revision of particle acceleration models within these objects. Prior campaigns with VERITAS have searched for TeV emission from various systems, including during two giant outbursts of the BeXRB 1A\,0535+262 in 2009 \cite{1A0535_2009} and 2020 \cite{1A0535_2020}, and during an outburst of 4U\,0115+634 in 2015 \cite{4U0115}. Here we report on recent VERITAS observations of another BeXRB system, LS V +44 17, during its first known giant outburst in early 2023.

\section{LS V +44 17}
LS V +44 17 was first identified as a likely massive X-ray binary (RX\,J0440.9+443) in a cross-correlation of the ROSAT Galactic Plane Survey with OB-star catalogues \cite{discovered}. It was subsequently confirmed as an accreting Be/X-ray binary system following the RXTE discovery of X-ray pulsations, with a period of $202.5\pm0.5\U{s}$ \cite{identified}. The orbital period is $150.0\pm0.2\U{days}$ \cite{Ferrigno2013}, determined from Swift/BAT observations of regular Type I outbursts around periastron. The system is located at a distance of $3.2^{+0.5}_{-0.6} \U{kpc}$ \cite{distance} and the massive star is classified as B0.2Ve \cite{stellartype}. The first evidence for transient X-ray behaviour was the observation of a Type I outburst discovered with MAXI/GSC in April 2010 \cite{firstoutburst}. Continuous X-ray monitoring since then revealed only occasional Type I outbursts (e.g. \cite{GBM}), until December 2022 when MAXI/GSC alerted the community to a dramatic X-ray brightening \cite{firstATEL}. After an initial peak and decline, the X-ray flux rebrightened until the start of February 2023, reaching more than twice the Crab flux in the hard X-ray band (Swift/BAT (15-50 keV) \cite{SwiftPeak}) before declining. These results are discussed in more detail later in these proceedings.

At higher energies, in the gamma-ray band, there are no nearby sources in the Fermi-LAT fourth source catalog in the energy range from 50 MeV to 1 TeV \cite{4FGL}. From the ground, historical observations were conducted by VERITAS as part of a binary system discovery program with a total exposure on LS~V~+44~17 of  $14.2\U{hours}$ collected between 2011 and 2016. No evidence for emission was found and the 99\% confidence level upper limit was $3.1\times10^{-13}\UU{cm}{-2}\UU{s}{-1}$ above $350\U{GeV}$ \cite{pastVTS}.

\section{VERITAS Observations during the 2023 outburst}

VERITAS is an array of four imaging atmospheric Cherenkov telescopes located at the Fred Lawrence Whipple Observatory in southern Arizona. The array is sensitive to gamma rays with energies between $\sim100\U{GeV}$ and $~30\U{TeV}$. In its current configuration, VERITAS is able to detect a source with 1\% of the Crab Nebula flux in $<25\U{hours}$ \cite{Park}.

At a declination of $+45^{\circ}$, and easily visible from September to March, the 2023 giant outburst of LS~V~+44~17 was well-situated for VERITAS follow-up. A first series of observations of LS~V~+44~17 with VERITAS began on January 24th, 2023 (MJD 59968) and continued nightly until January 27th. Observations were taken in the standard \textit{wobble} mode, in which the source is offset from the center of the field-of-view by $0.5^{\circ}$ to allow for background estimation.  The total exposure during this period, after corrections to remove data affected by poor weather or hardware problems, was $10.5\U{hours}$.  No evidence for emission was found, and the results were distributed promptly by astronomer's telegram \cite{VERITAS_ATEL}. Although the X-ray flux had not yet peaked, further gamma-ray observations with VERITAS were not immediately possible due to poor weather and the full Moon (atmospheric Cherenkov telescopes require clear and moderately dark skies to operate). A second series of observations were made shortly after the X-ray peak, on February 10th (MJD 59985), totalling $1.9\U{hours}$. The nightly exposures are listed in Table~\ref{tab1}.

\begin{table}
\centering
\begin{tabular}{|c|c|c|c|c|c|} 
\hline  Date & Start time (MJD) & Stop time (MJD) & Duration (hours)\\ 
\hline  2023-01-24 & 59968.0931 & 59968.1146 & 0.4 \\
\hline 2023-01-25 & 59969.0896 & 59969.2660 & 4.0 \\
\hline 2023-01-26 & 59970.0903 & 59970.2917 & 4.7 \\
\hline 2023-01-27 &  59971.2250 & 59971.2875 & 1.4 \\
\hline 2023-02-10 & 59985.0993 & 59985.2028 & 1.9 \\
\hline 
\end{tabular}
\caption{Summary of VERITAS observations of LS V +44 17 during the giant outburst in 2023}
\label{tab1}
\end{table}

\begin{figure}[ht]
    \centering
    \includegraphics[width=\textwidth]{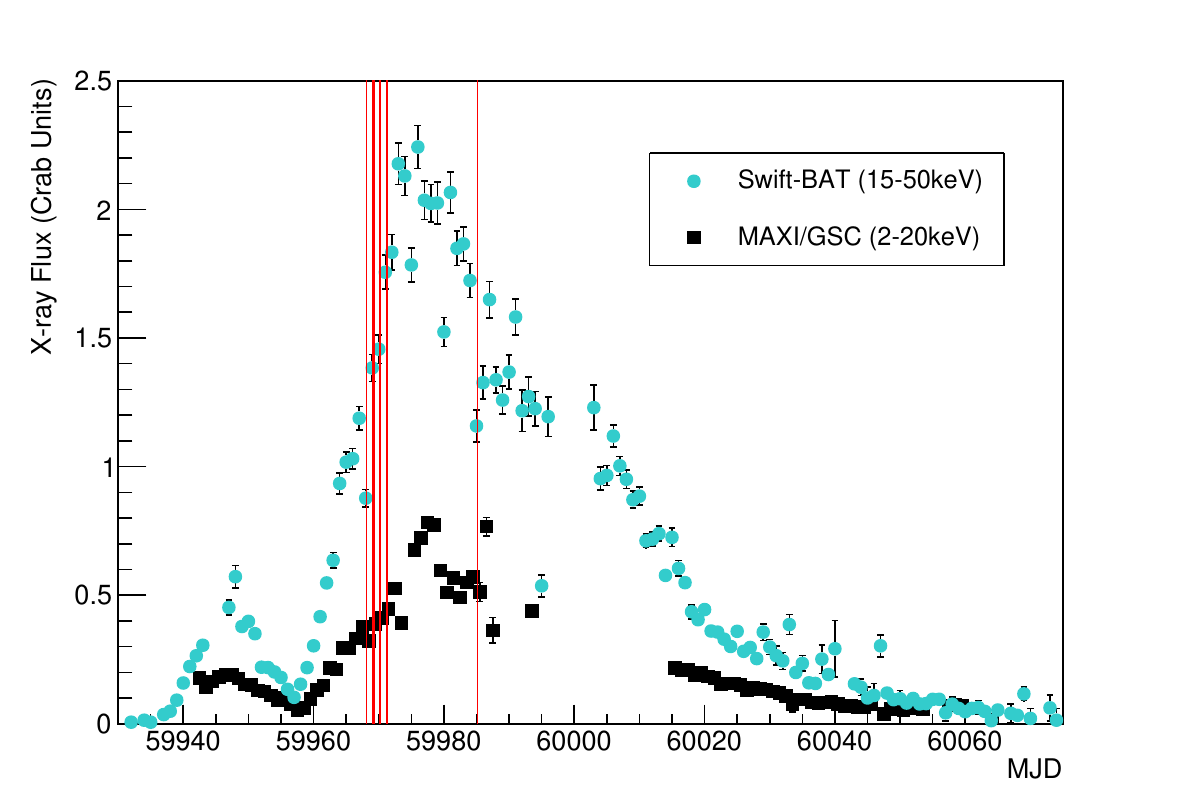}
    \captionof{figure}{X-ray lightcurves of LS~V~+44~17 during the giant outburst in 2023. Red vertical lines indicate the times of VERITAS gamma-ray observations.}
    \label{Xrays}
\end{figure}

Figure~\ref{Xrays} shows the X-ray light curves from Swift-BAT ($15-50\U{keV}$)\footnote{\url{https://swift.gsfc.nasa.gov/results/transients/weak/LSVp4417/}} \cite{SwiftMonitor} and MAXI/GSC\footnote{\url{http://maxi.riken.jp/pubdata/v7.7l/J0440+445/index.html}} ($2-20\U{keV}$) during the entire outburst, with the VERITAS observing periods indicated. The VERITAS exposures sample both the rising and falling edges of the flare, in both cases when the X-ray flux was approximately 75\% of the peak value.

We have re-analyzed the full 2023 VERITAS dataset here. Observations were processed using standard VERITAS analysis tools which parameterize images of the Cherenkov light from air showers in order to discriminate gamma-ray events from the cosmic ray background, and to reconstruct the arrival direction and energy of the primary photons \cite{evndisp}. There is no evidence for emission in the total dataset, nor on any of the individual days. The upper limit is $2.1\times10^{-12}\UU{cm}{-2}\UU{s}{-1}$ above $200\U{GeV}$ at 99\% confidence for an assumed power-law with an index of -2.4.

\section{Discussion}

\begin{figure}[ht]
    \centering
    \includegraphics[width=\textwidth]{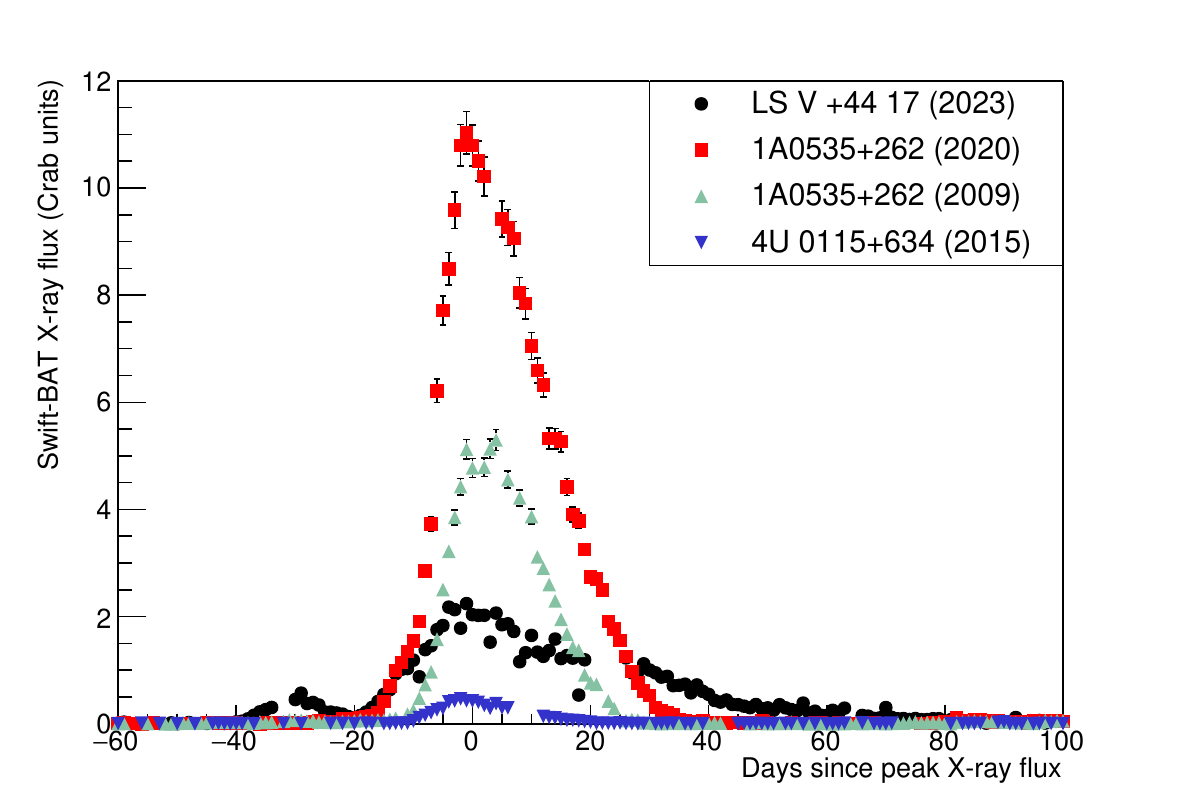}
    \captionof{figure}{Swift-BAT X-ray lightcurves of all of the Type II outbursts from BeXRBs observed by VERITAS. Data are taken from the Swift/BAT Hard X-ray Transient Monitoring website \cite{SwiftMonitor}.}
    \label{allbursts}
\end{figure}

Figure~\ref{allbursts} places our results in context with all other Type II outbursts from BeXRBs which have been observed with VERITAS. In X-ray flux, the outburst from LS V +44 17 lies between 4U\,0115+634 and the two bursts seen from 1A\,0535+262, which is primarily a consequence of their distances: 1A\,0535+262 is among the closest XRBs, at a distance of $\sim2\U{kpc}$, while LS~V~+44~17 and 
4U\,0115+634 are at $3.2^{+0.5}_{-0.6} \U{kpc}$ and $7.2^{+1.5}_{-1.1} \U{kpc}$, respectively \cite{GaiaDR2_distance}. The upper limits to the gamma-ray luminosity are typically a few percent of the X-ray luminosity and a tiny fraction of the Eddington luminosity suggesting that, unlike their pulsar-wind driven counterparts, accreting X-ray pulsars are not efficient at accelerating particles to high energies, and are not promising targets for future ground-based gamma-ray observatories. 

\section{Acknowledgements}
This research is supported by grants from the U.S. Department of Energy Office of Science, the U.S. National Science Foundation and the Smithsonian Institution, by NSERC in Canada, and by the Helmholtz Association in Germany. This research used resources provided by the Open Science Grid, which is supported by the National Science Foundation and the U.S. Department of Energy's Office of Science, and resources of the National Energy Research Scientific Computing Center (NERSC), a U.S. Department of Energy Office of Science User Facility operated under Contract No. DE-AC02-05CH11231. We gratefully acknowledge the excellent work of the technical support staff at the Fred Lawrence Whipple Observatory and at the collaborating institutions in the construction and operation of the instrument.

\clearpage

\section*{Full Author List: VERITAS Collaboration}

\scriptsize
\noindent
A.~Acharyya$^{1}$,
C.~B.~Adams$^{2}$,
A.~Archer$^{3}$,
P.~Bangale$^{4}$,
J.~T.~Bartkoske$^{5}$,
P.~Batista$^{6}$,
W.~Benbow$^{7}$,
J.~L.~Christiansen$^{8}$,
A.~J.~Chromey$^{7}$,
A.~Duerr$^{5}$,
M.~Errando$^{9}$,
Q.~Feng$^{7}$,
G.~M.~Foote$^{4}$,
L.~Fortson$^{10}$,
A.~Furniss$^{11, 12}$,
W.~Hanlon$^{7}$,
O.~Hervet$^{12}$,
C.~E.~Hinrichs$^{7,13}$,
J.~Hoang$^{12}$,
J.~Holder$^{4}$,
Z.~Hughes$^{9}$,
T.~B.~Humensky$^{14,15}$,
W.~Jin$^{1}$,
M.~N.~Johnson$^{12}$,
M.~Kertzman$^{3}$,
M.~Kherlakian$^{6}$,
D.~Kieda$^{5}$,
T.~K.~Kleiner$^{6}$,
N.~Korzoun$^{4}$,
S.~Kumar$^{14}$,
M.~J.~Lang$^{16}$,
M.~Lundy$^{17}$,
G.~Maier$^{6}$,
C.~E~McGrath$^{18}$,
M.~J.~Millard$^{19}$,
C.~L.~Mooney$^{4}$,
P.~Moriarty$^{16}$,
R.~Mukherjee$^{20}$,
S.~O'Brien$^{17,21}$,
R.~A.~Ong$^{22}$,
N.~Park$^{23}$,
C.~Poggemann$^{8}$,
M.~Pohl$^{24,6}$,
E.~Pueschel$^{6}$,
J.~Quinn$^{18}$,
P.~L.~Rabinowitz$^{9}$,
K.~Ragan$^{17}$,
P.~T.~Reynolds$^{25}$,
D.~Ribeiro$^{10}$,
E.~Roache$^{7}$,
J.~L.~Ryan$^{22}$,
I.~Sadeh$^{6}$,
L.~Saha$^{7}$,
M.~Santander$^{1}$,
G.~H.~Sembroski$^{26}$,
R.~Shang$^{20}$,
M.~Splettstoesser$^{12}$,
A.~K.~Talluri$^{10}$,
J.~V.~Tucci$^{27}$,
V.~V.~Vassiliev$^{22}$,
A.~Weinstein$^{28}$,
D.~A.~Williams$^{12}$,
S.~L.~Wong$^{17}$,
and
J.~Woo$^{29}$\\
\\
\noindent
$^{1}${Department of Physics and Astronomy, University of Alabama, Tuscaloosa, AL 35487, USA}

\noindent
$^{2}${Physics Department, Columbia University, New York, NY 10027, USA}

\noindent
$^{3}${Department of Physics and Astronomy, DePauw University, Greencastle, IN 46135-0037, USA}

\noindent
$^{4}${Department of Physics and Astronomy and the Bartol Research Institute, University of Delaware, Newark, DE 19716, USA}

\noindent
$^{5}${Department of Physics and Astronomy, University of Utah, Salt Lake City, UT 84112, USA}

\noindent
$^{6}${DESY, Platanenallee 6, 15738 Zeuthen, Germany}

\noindent
$^{7}${Center for Astrophysics $|$ Harvard \& Smithsonian, Cambridge, MA 02138, USA}

\noindent
$^{8}${Physics Department, California Polytechnic State University, San Luis Obispo, CA 94307, USA}

\noindent
$^{9}${Department of Physics, Washington University, St. Louis, MO 63130, USA}

\noindent
$^{10}${School of Physics and Astronomy, University of Minnesota, Minneapolis, MN 55455, USA}

\noindent
$^{11}${Department of Physics, California State University - East Bay, Hayward, CA 94542, USA}

\noindent
$^{12}${Santa Cruz Institute for Particle Physics and Department of Physics, University of California, Santa Cruz, CA 95064, USA}

\noindent
$^{13}${Department of Physics and Astronomy, Dartmouth College, 6127 Wilder Laboratory, Hanover, NH 03755 USA}

\noindent
$^{14}${Department of Physics, University of Maryland, College Park, MD, USA }

\noindent
$^{15}${NASA GSFC, Greenbelt, MD 20771, USA}

\noindent
$^{16}${School of Natural Sciences, University of Galway, University Road, Galway, H91 TK33, Ireland}

\noindent
$^{17}${Physics Department, McGill University, Montreal, QC H3A 2T8, Canada}

\noindent
$^{18}${School of Physics, University College Dublin, Belfield, Dublin 4, Ireland}

\noindent
$^{19}${Department of Physics and Astronomy, University of Iowa, Van Allen Hall, Iowa City, IA 52242, USA}

\noindent
$^{20}${Department of Physics and Astronomy, Barnard College, Columbia University, NY 10027, USA}

\noindent
$^{21}${ Arthur B. McDonald Canadian Astroparticle Physics Research Institute, 64 Bader Lane, Queen's University, Kingston, ON Canada, K7L 3N6}

\noindent
$^{22}${Department of Physics and Astronomy, University of California, Los Angeles, CA 90095, USA}

\noindent
$^{23}${Department of Physics, Engineering Physics and Astronomy, Queen's University, Kingston, ON K7L 3N6, Canada}

\noindent
$^{24}${Institute of Physics and Astronomy, University of Potsdam, 14476 Potsdam-Golm, Germany}

\noindent
$^{25}${Department of Physical Sciences, Munster Technological University, Bishopstown, Cork, T12 P928, Ireland}

\noindent
$^{26}${Department of Physics and Astronomy, Purdue University, West Lafayette, IN 47907, USA}

\noindent
$^{27}${Department of Physics, Indiana University-Purdue University Indianapolis, Indianapolis, IN 46202, USA}

\noindent
$^{28}${Department of Physics and Astronomy, Iowa State University, Ames, IA 50011, USA}

\noindent
$^{29}${Columbia Astrophysics Laboratory, Columbia University, New York, NY 10027, USA}

%

\end{document}